\documentclass[5p,twocolumn]{elsarticle}
\usepackage{graphicx,latexsym}
\usepackage{dcolumn}
\usepackage{amssymb,amsmath,bm}
\usepackage{subfigure}
\usepackage{braket}
\usepackage{siunitx}
\usepackage{array}
\usepackage{float}
\usepackage[nopar]{lipsum}
\usepackage{caption}
\usepackage{multirow}
\usepackage{bigstrut}
\usepackage[super]{nth}

\newcommand{\angstrom}{\text{\normalfont\AA}}
\usepackage{hyperref}
\hypersetup{
    pdfnewwindow=true,       % links in new window
    colorlinks=true,         % false: boxed links; true: colored links
    linkcolor=blue,          % color of internal links
    citecolor=blue,          % color of links to bibliography
    filecolor=magenta,       % color of file links
    urlcolor=black           % color of external links
}

\usepackage[normalem]{ulem}

\def\sec#1{Sec.\ \ref{#1}}
\def\eq#1{Eq.\ (\ref{#1})}
\def\fig#1{Fig.\ \ref{#1}}

\journal{}

\begin{document}

\begin{frontmatter}

%-----------------------------------------------------------------

\title{High thermoelectric and optical conductivity driven by the interaction of Boron and Nitrogen dopant atoms with a 2D monolayer Beryllium Oxide}

\author[a1,a2]{Nzar Rauf Abdullah}
\ead{nzar.r.abdullah@gmail.com}
\address[a1]{Division of Computational Nanoscience, Physics Department, College of Science, 
             University of Sulaimani, Sulaimani 46001, Kurdistan Region, Iraq}
\address[a2]{Computer Engineering Department, College of Engineering, Komar University of Science and Technology, Sulaimani 46001, Kurdistan Region, Iraq}

\author[a3]{Botan Jawdat Abdullah}
\address[a3]{Physics Department, College of Science, Salahaddin University-Erbil, Erbil, Kurdistan Region, Iraq}

\author[a4]{Vidar Gudmundsson}
\address[a4]{Science Institute, University of Iceland, Dunhaga 3, IS-107 Reykjavik, Iceland}

%----------------------------------------------------------------

\begin{abstract}

The electronic, thermal and optical properties of a monolayer BeO with Boron (B) and Nitrogen (N) co-dopant atoms are studied by means of a density functional theory computation. 
Our calculations reveal that BeO with BN-codopant atoms can give rise to more effective and outstanding performance for the thermal and optical responses. More significantly, the monolayer BeO with BN codopant atoms becomes a semiconductor with a direct band gap in comparison with the insulator behavior of pristine BeO. The particular attention of this work is paid to the influence of the atomic configuration and the interaction of the B and N dopant atoms with BeO. The interaction of the B and N atoms with the BeO monolayer diminishes degenerate energy states forming flat bands.
It is also found that there is a strong attractive interaction between the O and N atoms forming a strong sigma bond breaking the symmetry of BeO structure. Consequently, the band gap is reduced leading to a semiconductor behavior with improved thermoelectric properties such as the Seebeck coefficient and the figure of merit. The reduced band gap and the flat bands induce a high optical responses such as the refractive index, the reflectivity and the optical conductivity in the visible light region. In addition, the anisotropy of a monolayer BeO with B and N atoms regarding different direction of electromagnetic polarization is presented. We anticipate that our results can be useful for design of both thermoelectric and optoelectronic devices.

\end{abstract}


\end{frontmatter}

\section{Introduction} 

Since the discovery of graphene and its first synthesis in 2004 \cite{Novoselov666}, significant research efforts have been directed toward the synthesis of other members of a broad family of two-dimensional (2D) materials. In addition, modern computational methods have been used to investigate newly undiscovered 2D materials and some of them have been synthesized in experiments \cite{doi:10.1021/nl402150r, PhysRevLett.108.155501, ABDULLAH2020126807, doi:10.1021/nn4009406, PhysRevLett.105.136805}. The 2D materials have thus attracted the attention of researchers due to their unique electrical, optical, thermal, mechanical, and magnetic properties, making them great candidates for a wide range of applications \cite{Mak2016, Mannix2018, ABDULLAH2020100740, Zhu2015, Ji2016}.

An interesting 2D material is a beryllium oxide (BeO) monolayer. 
Experimentally, a thin BeO was initially synthesized by oxidizing a cleansed Be surface naturally in 2009 \cite{Reinelt_2009}. Monolayer BeO is a new graphene-like monoxide structure that has recently been synthesized using molecular beam epitaxy of a single atomic sheet with a honeycomb pattern on an Ag thin film \cite{doi:10.1021/acsnano.0c06596}. The physical characteristics of an atomic layer of BeO grown on Si and GaAs substrates have been investigated as a barrier/passivation layer in III-V devices. Compared to Al$_2$O$_3$, BeO shows a superior self-cleaning effect, a reduced interface defect density, frequency dispersion, leakage current density, and hysteresis \cite{doi:10.1063/1.3553872}. 

The combined experimental and theoretical research also indicates the formation of a honeycomb structure of BeO monolayer on the Mo(112) surfaces \cite{TV2018}. 
On the other hand, the physical characteristics of a Ge/BeO heterostructure have been investigated using density functional theory, DFT, with the findings indicating that the BeO monolayer is an ideal substrate for germanene \cite{doi:10.1021/acs.jpcc.6b06161}, and structural stability and electrical characteristics of the Sn/BeO heterostructure have been investigated through DFT. The results demonstrate that in a BeO monolayer the band gap opens significantly, while also retaining sufficiently different electrical characteristics of stanene to a great extent \cite{Chakraborty_2020}.

This new monolayer has been theoretically investigated using first principles calculations. For instance, the monolayer BeO exhibits high dynamic stability due to atomic hybridization and electronic delocalization \cite{PhysRevB.92.115307, Luo17213}. The optical conductivity in a parallel ($E\parallel x$) or perpendicular ($E\perp z$) polarized electric field to the BeO nanosheet indicates the existence of a semiconductor energy gap for $E\parallel x$ and an insulator for $E\perp z$. The optical spectra are anisotropic, and the static refractive index of the nanosheet structure is lower than that of the wurtzite structure \cite{Valedbagi2014}. The monolayer h-BeO exhibits a wide band gap due to the atomic electronegativity difference in the polar h-BeO. The biaxial tensile strain decreases electron effective mass and therefore increases the electron-phonon coupling strength as well, as the electrical transport is restricted by phonons \cite{doi:10.1063/5.0022426}.

Furthermore, electrons in a monolayer BeO are found to be strongly concentrated around the O and Be atoms, which differs from graphene indicating an ionic character of the Be-O bond. The thermal conductivity of a planar BeO monolayer reduces as the temperature rises, which is mostly due to low frequency phonons \cite{Xia_2020}. If the planar buckling parameter is increased, a decrease in the energy band gap occurs. This shows that the planar buckling can remove energy level degeneracy in the conduction band, and the optical spectra are red-shifted to lower energy by raising the planar buckling factor \cite{JALILIAN2016120}. Therefore, the BeO monolayer is expected to have a high electronic band gap, elastic modulus, tensile strength, and lattice thermal conductivity through DFT calculations.

Doping is the most common practical approach for controlling the properties of effective materials and essentially modifying the electrical, thermal, optical, and magnetic properties of host 2D-layered materials, as well as altering them depending on the applications \cite{nano11040832, doi:10.1021/acs.nanolett.5b00314, zhang_robinson_2019}. The dopants have a significant effect on the behavior of the 2D monolayer materials. 
Since the BeO monolayer is almost a new material to study, there has only been a limited work on it via doping. DFT has been utilized in some theoretical studies to examine the electrical and magnetic properties of monolayer BeO with transition metal (TM) substitutional doping. The results show that the electrical and magnetic properties of monolayer BeO with various TM substitutional doped BeO systems can be tuned \cite{SONG2018252}. The magnetic behavior of an N-doped BeO monolayer is produced by spin-up and spin-down band gaps, which vary depending on the dopant concentration and the N atom spacing \cite{Hoat_2021}.

The monolayer BeO has olny been minimaly investigated via doping, despite the fact that it is extremely desirable for identifying unique characteristics. Therefore, the electrical, the thermal, and the optical characteristics of a monolayer BeO codoped with B and N (Be$_x$O$_x$B$_{1-x}$N$_{1-x}$) are investigated using first-principles calculations in this study. The results demonstrate that the B and N codopants have a significant effect on the BeO monolayer, causing the band structures to alter, lowering the energy gap while preserving the structure as a semiconductor. As a consequence, the obtained results demonstrate that the electrical, the thermal, and the optical properties are enhanced. It indicates that the B and N codoped BeO monolayer may play an important role for thermoelectric and optoelectronic applications.

In \sec{Sec:Computational} the computational techniques and the model structure are briefly overviewed. In \sec{Sec:Results} the main achieved results are analyzed. In \sec{Sec:Conclusion} the conclusion of results is presented.

\section{Computational Tools}\label{Sec:Computational}

First, the software used in this study for crystalline and molecular structure visualization are (XCrySDen) and VESTA, which can be used to visualize the BeO and BN-codoped BeO monolayers \cite{KOKALJ1999176, momma2011vesta}. All the electronic and optical calculations are performed by using the density functional theory (DFT) as implemented in the Quantum Espresso, QE, code \cite{Giannozzi_2009, giannozzi2017advanced}. 
The generalized gradient approximation (GGA) in the Perdew-Burke- Ernzernhof (PBE) functionals is utilized to investigate the electronic band structure, the density of states, DOS, and the optical properties \cite{PhysRevLett.77.3865, doi:10.1063/1.1926272}.

The relaxation parameters for the fully relaxed monolayers are important to highlight. We use an  
$18\times18\times1$ Monkhorst-Pack K-point grid and an energy cutoff $1088$~eV. Furthermore, 
the atomic positions are considered fully relaxed when the forces are smaller than $0.001$ eV/$\angstrom$. 
A vacuum space of $20 \, \angstrom$ is introduced to ensure there is no interaction between BeO sheets.

After the full relaxation of all monolayers on a K-point $18\times18\times1$-grid for static self-consistency we
use a $100\times100\times1$ grid for the DOS calculations with the same 
energy cutoff as is used in the relaxation. An optical broadening of $0.1$~eV is assumed for calculating the dielectric properties. Finally, to calculate the thermometric properties and the Boltzmann transport properties the software package (BoltzTraP) is utilized \cite{Madsen2006, ABDULLAH2021106073}. The BoltzTraP code uses a mesh of band energies and has an interface to the QE package \cite{ABDULLAH2021106981, ABDULLAH2021110095}.

\section{Results}\label{Sec:Results}

The structures under investigation are pure and BN-codoped BeO monolayers with 
$2\times2\times1$ supercell. The electronic, optical, and thermoelectric properties are 
the main topics of this work.

\subsection{Electronic properties}

\begin{figure}[htb]
	\centering
	\includegraphics[width=0.22\textwidth]{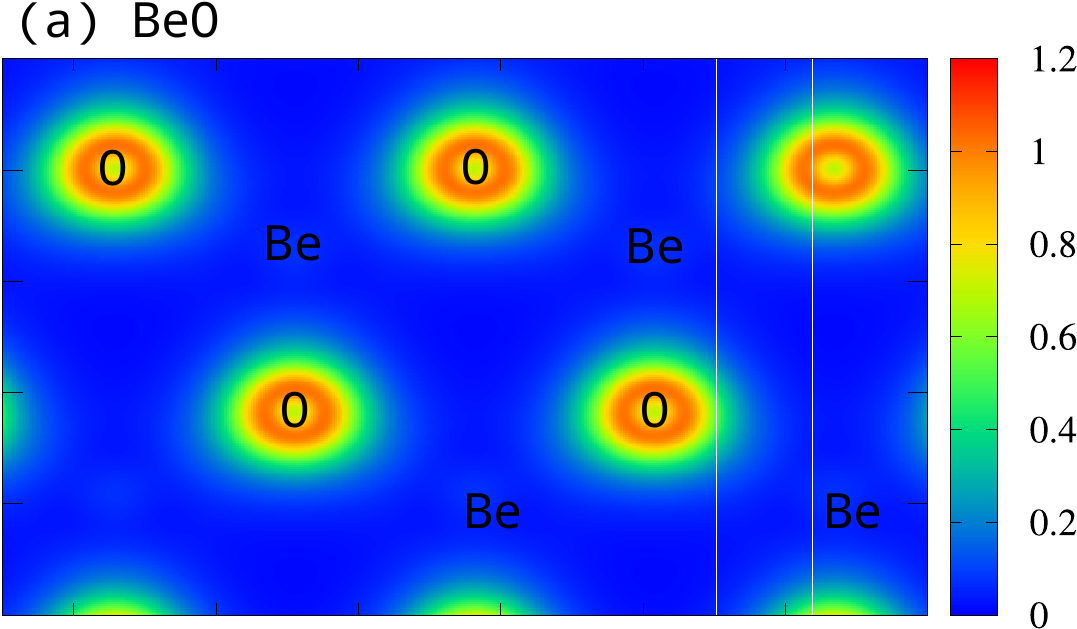}
	\includegraphics[width=0.22\textwidth]{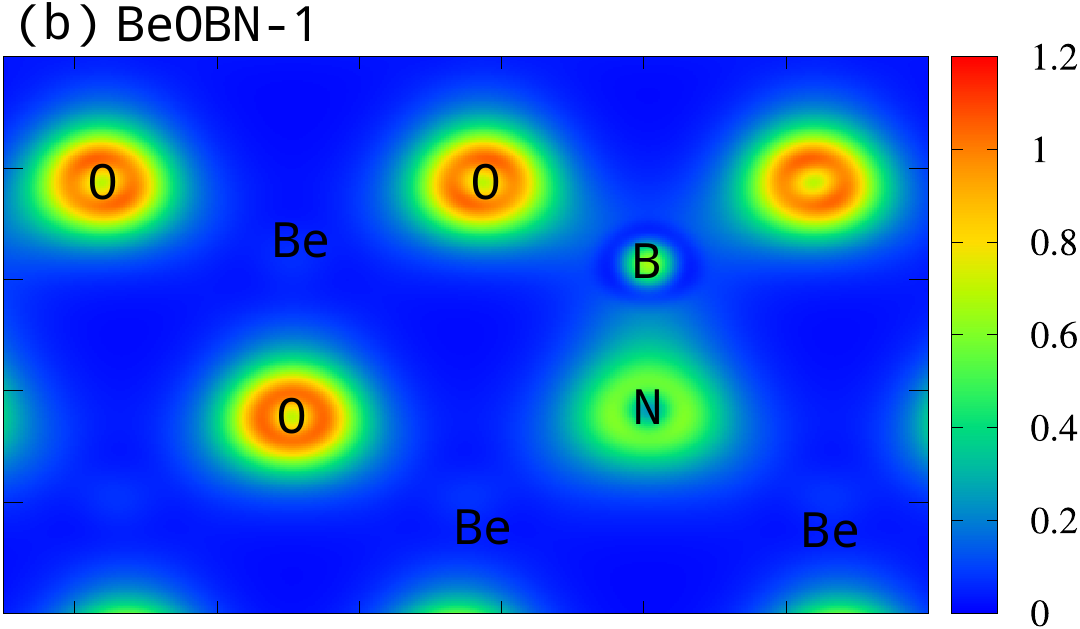}\\
	\includegraphics[width=0.22\textwidth]{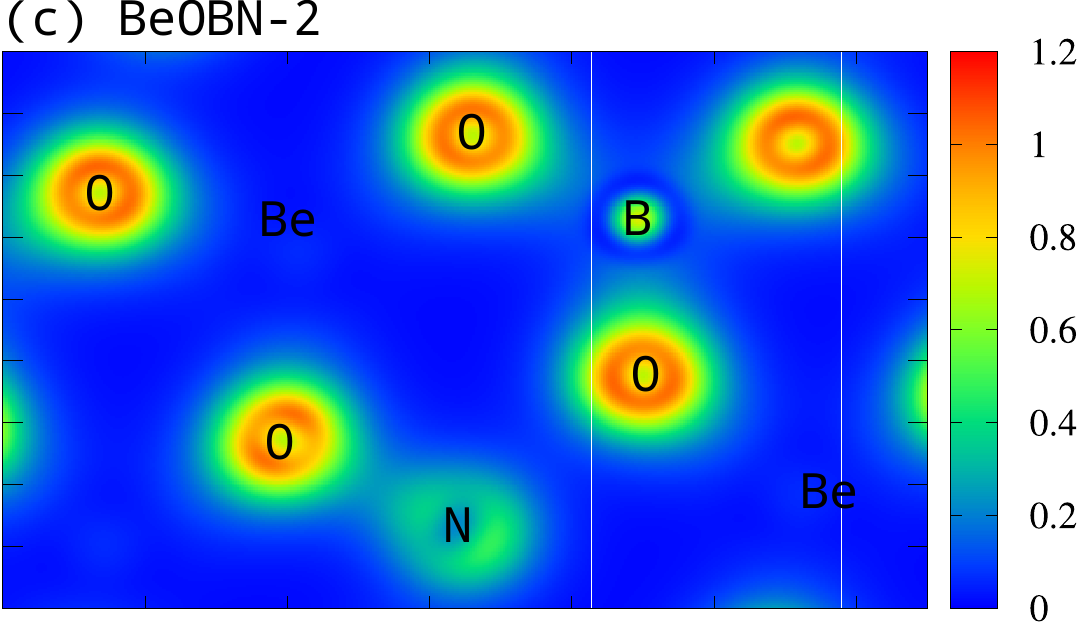}
	\includegraphics[width=0.22\textwidth]{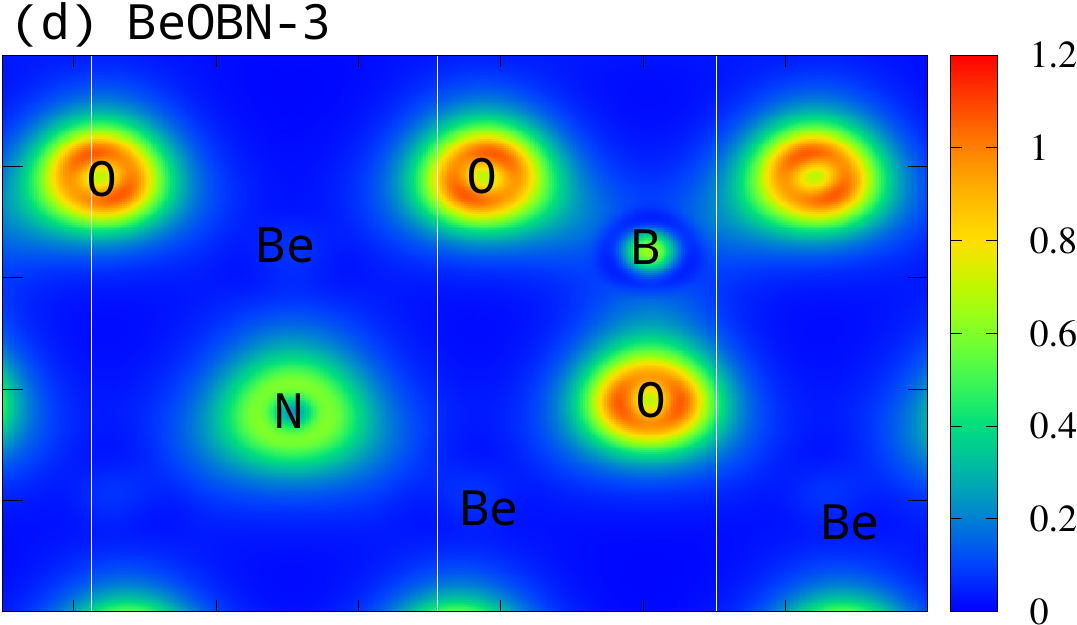}
	\caption{Electron charge distribution of BeO (a), BeOBN-1 (b), BeOBN-2 (c), and BeOBN-3 (d).}
	\label{fig01}
\end{figure}

The positions of the atoms in the hexagonal monolayer $2\times2$ supercell are the para, the ortho, and the meta position \cite{RASHID2019102625}.
Three atomic configurations for the B and N dopant atoms in BeO are thus possible when a BN-copdoped BeO monolayer is considered. We show these three possible atomic configurations in \fig{fig01} together with their electron charge distribution.
The electron charge distribution of pure BeO monolayer is first displayed in \fig{fig01}(a), and we find that the electron charge distribution around the Oxygen atoms is much higher than that of the Beryllium atoms. This is expected as the electronegativity of the Oxygen atom, $3.44$, is much higher than that of the Beryllium atom, $1.57$.

Back to the three possible configurations of B and N atoms in BeO. First, the B and N atoms are doped at the ortho- and meta-position, respectively, identified as the BeOBN-1 monolayer (see \fig{fig01}(b)) \cite{ABDULLAH2020103282}. Second, the B atom is kept at the ortho-position but the N atom is doped at the para-position identified as BeOBN-2 (c). Third, the B atom is still at the ortho-position while 
the N atom is doped at the meta-position but in the opposite half of the hexagon (d). 
These three atomic configurations will gives rise different physical characteristics for the BN-codoped BeO monolayer as the interaction between atoms in these three configurations are different.

In all the three cases for BeOBN monolayers, a strong electron charge distribution around the N atoms is seen, while a weak electron charge distribution around the B atom is found. The strong charge localization of the N atoms is casued by its electronegativity, $3.04$, which is higher than that for B atoms, $2.04$, and lower than the electronegativity of the O atoms.

In the atomic configuration for the BeOBN-1 monolayer, a B atom is only bonded to  O and N atoms, while an N atom is only connected to B and Be atoms. Consequently, the B and the N atoms prefer to form $\sigma$-bonds (B-N bonds) as their electronegativity is close to each other. The electron charge distribution between the B and N atoms confirms the B-N $\sigma$-bond.

In the atomic configuration of a BeOBN-2 monolayer, both the B and N atoms are located between O atoms and no direct connection between B(N) and Be atoms is seen. As the electronegativities of N and O are very close, a strong attractive interaction between the N and O atoms is observed. As a result, there is an electron transfer from the N atom to the O atom. Electron charge distribution in the N-O bond is clearly formed. Furthermore, a weak tail of charge between the B and O atoms is formed indicating a weak attractive interaction between these atoms in the structure. 

In the last atomic configuration, in the BeOBN-3 monolayer, a B atom is only connected to an O atom forming a B-O bond while an N atom is only bonded to  Be atoms forming N-Be bonds. One can see that there is no electron charge distribution between the N and the Be atoms indicating no interaction between the N and Be atoms, while a weak electron charge distribution between the B and O atoms is observed.
The atomic configuration of the BeOBN-1 and the BeOBN-3 monolayers are very similar except that in the  BeOBN-1 B-N bonds are formed. 

The energetic stability of BeOBN monolayers can be realized from a calculation of their formation energy.
The formation energy is the energy needed for forming the atomic configuration of a structure \cite{ABDULLAH2020114556}. The DFT calculations show that the three BeOBN monolayers can be arranged from the higher to the lower formation energy as follow: BeOBN-2 $<$ BeOBN-1 $<$ BeOBN-3.
It has been shown that the lower the formation energy, the more energetically stable structure is obtained. Consequently, BeOBN-2 layers are the most stable structures. 
We note that the structure is more stable if the dopant atoms are put at the Be positions of the hexagon of the BeO monolayers.

\begin{figure}[htb]
	\centering
	\includegraphics[width=0.45\textwidth]{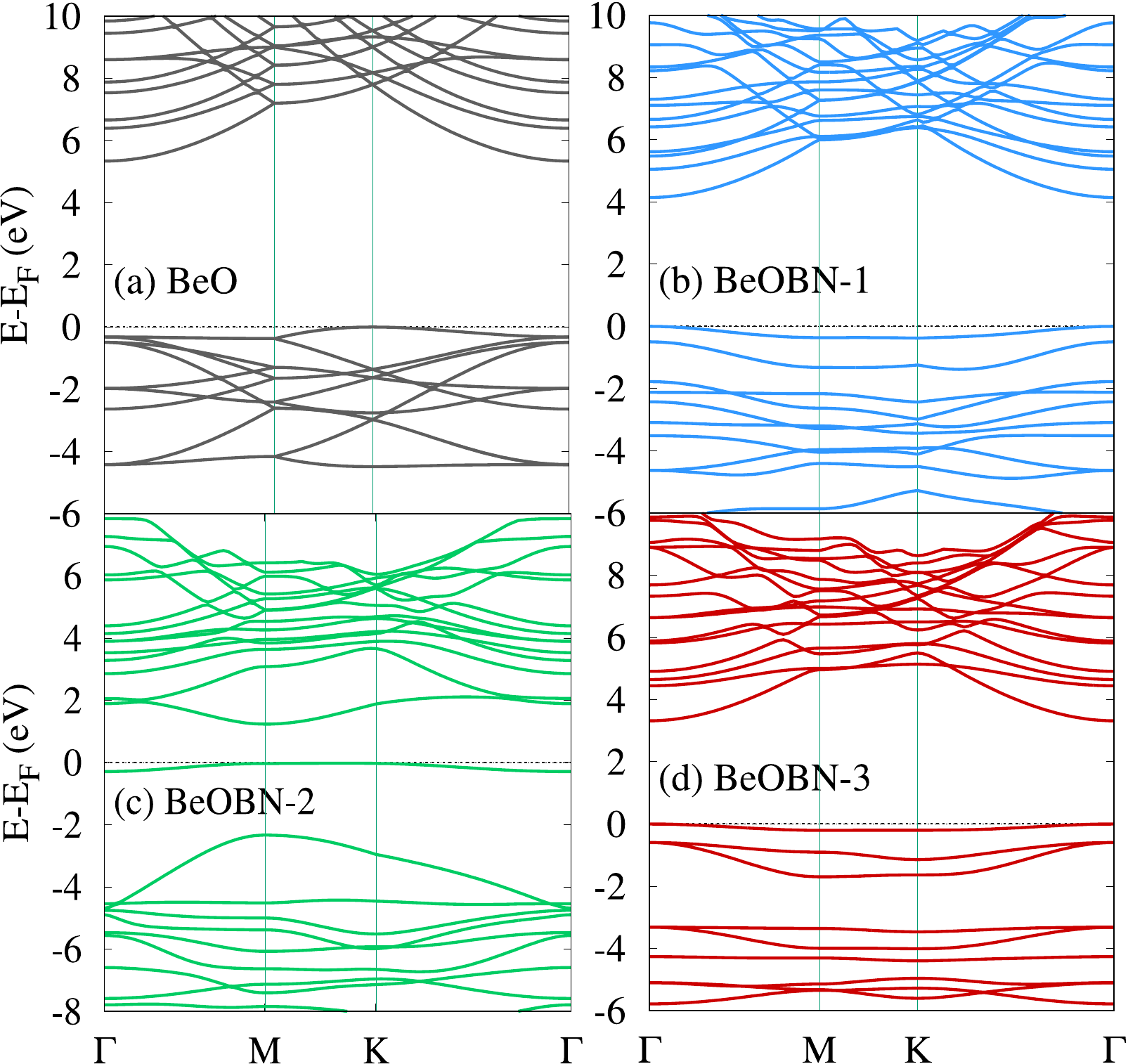}
	\caption{Band structure for optimized structures of BeO (a), BeOBN-1 (b), BeOBN-2 (c), and BeOBN-3 (d). The energies are with respect to the Fermi level, and the Fermi energy is set to zero.}
	\label{fig02}
\end{figure}

In order to see effects of the atomic configurations on the energy dispersions, we present the band structures in \fig{fig02} for the pure BeO monolayer (a) and the BeOBN monolayers (b-d).
The indirect band gap of a pure BeO monolayer is found to be $5.4$~eV displaced
from the K- to the $\Gamma$-point in our calculation, which is well in agreement with a recent PBE functional calculation \cite{MORTAZAVI2021100257}. The band gap of pure BeO is found to be $6.72$~eV using HSE06 functionals, and an experimental observation shows that the BeO monolayer is an insulator with a band gap of $6.4$~eV. To get insight to the band structure of pure BeO, we also present the partial density of states (PDOS) in \fig{fig03}(a). 
The PDOS of a pure BeO monolayer indicates that the valence bands are generated by a hybridization of $p$-orbitals of both the Be and O atoms, where the O atoms account for the main contributions. 
The conduction bands of BeO are mainly formed due to the $p$-orbital of the
Be atoms. The valence band maxima and the conduction band minima are mainly due to the $p_z$-orbitals of both the Be and O atoms, respectively (not shown).

In BeOBN monolayers, the band gap and the shape of the valence and the conduction bands are modified. 
It has been reported that the N dopant atoms shift the Fermi energy of BeO to the valence band region, while the B dopant atoms shift the Fermi energy to the conduction band region generating degenerate semiconductor properties \cite{abdullah2021enhanced}.
With the BN codopant atoms in a BeO layer, the semiconductor characteristics are seen without any crossing of the Fermi energy to either the valence or the conduction region as is seen in \fig{fig02}(b-d), and the pronounced effects of the atomic configuration of the B and N atoms on the band gap and the band structures are clearly seen as follows:  
\begin{figure}[htb]
	\centering
	\includegraphics[width=0.45\textwidth]{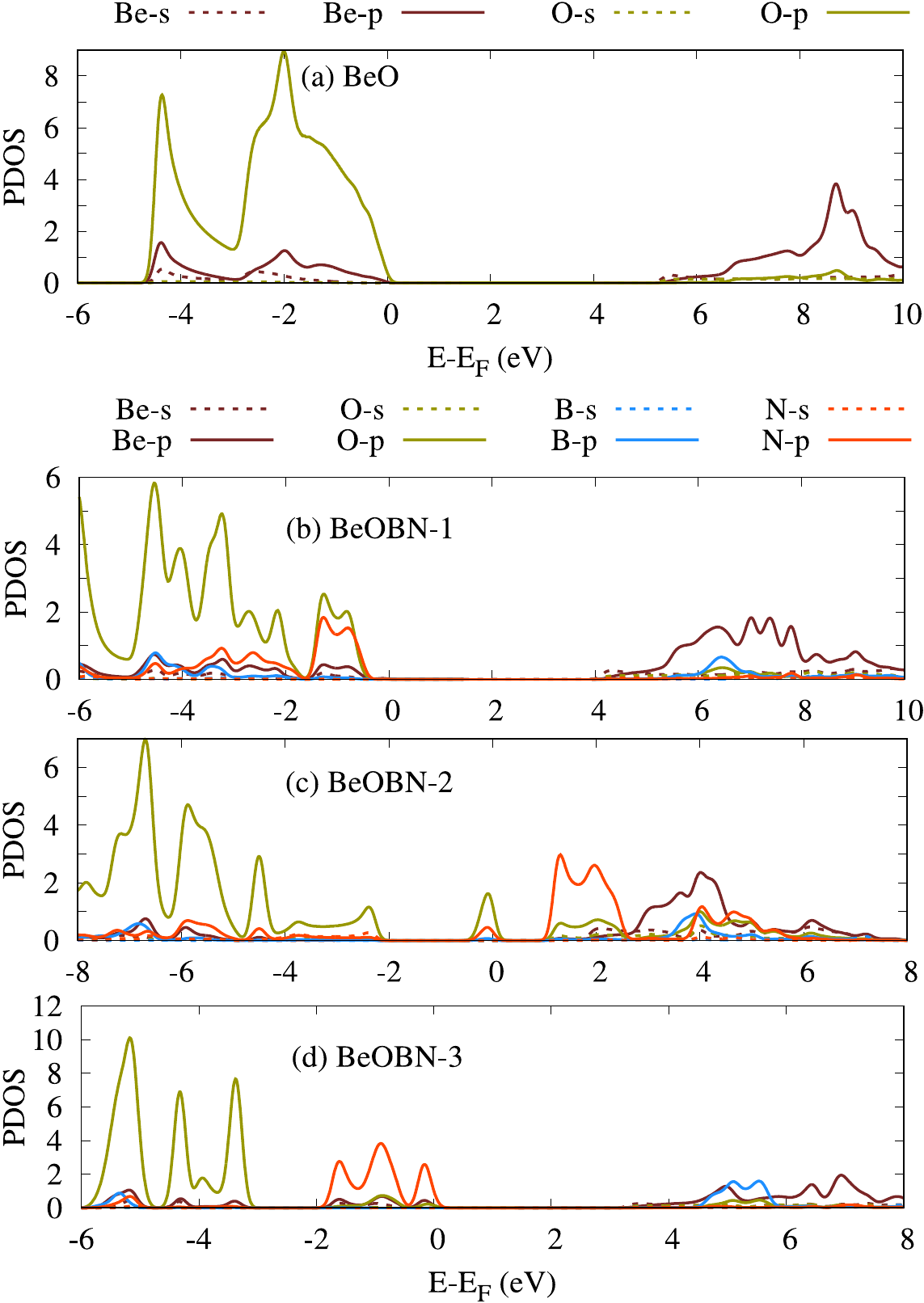}
	\caption{Partial density of states (PDOS) of pure BeO (a), BeOBN-1 (b), BeOBN-2 (c), and BeOBN-3 (d). The PDOS of s- (dashed lines) and the p-orbital (solid lines) of all four atoms (Be, O, B, and N) are plotted. The Fermi energy is set to zero.}
	\label{fig03}
\end{figure}
First, in the band structure of the BeOBN monolayers, several degeneracy points in the valence band region have vanished, and flat or parallel bands are formed. Second, the band gap is reduced as the conduction bands shift down to the lower part of energy. The reduction of the band gap depends on the atomic configuration of the B and N atoms. The band gap is $4.14$, $1.21$, $3.32$~eV for BeOBN-1, BeOBN-2, and BeOBN-3, respectively. The band structures of BeOBN-1 and BeOBN-3 have direct band gaps along the $\Gamma$ point while the BeOBN-2 has an indirect band gap. 

In order to better understand the aforementioned effects on the band structure, the PDOS of BeOBN monlayers is plotted in \fig{fig03}(b-d). In both BeOBN-1 and BeOBN-3 monolayers, a B(N) atom is doped at the position a Be(O) atom. Consequently, the new states are generated in the valence band region near to the Fermi energy due to the $p$-orbitals of N atoms (red solid line), and new states in the conduction band region are formed due to the $p$-orbitals of the B atoms (blue solid line). The contribution of the N and B atoms in PDOS is dominant giving a rise to a modulation in a high symmetry breaking. The degeneracy in valence band region is cancelled and the reduction of band gap is observed.

Most interesting is the band structure of BeOBN-2, where N atoms are doped at the Be atom positions, which is opposite to the BeOBN-1 and BeOBN-3 monolayers, where the N atom is put at the position of O atom. So, new states due to the N atom are generated in the conduction band region, which strongly shifts the conduction band minima down to energy reducing the band gap. Another reason for the strong modification of BeOBN-2 band structure are the atomic interactions. As we have seen from \fig{fig01}(c) for BeOBN-2, the strong interaction between the N-O atoms strongly influences a high symmetry breaking. As a result, the band gap is strongly reduced for the BeOBN-2 monolayer.

\subsection{Optical properties}

The parallel bands and the reduced band gap of BeOBN have a significant effects on their optical characteristics. To study the optical properties of the monolayers, we calculate the dielectric function, which demonstrates the linear response of the systems subjected to electromagnetic radiation.
In order to show the influence of the parallel bands, we present the imaginary part of the dielectric function estimating the absorption spectra, $\varepsilon_2$, the excitation spectra, $k$, the real part of dielectric function, $\varepsilon_1$, the refractive index, $n$, the reflectivity, and the optical conductivity.

\begin{figure*}[htb]
	\centering
	\includegraphics[width=0.85\textwidth]{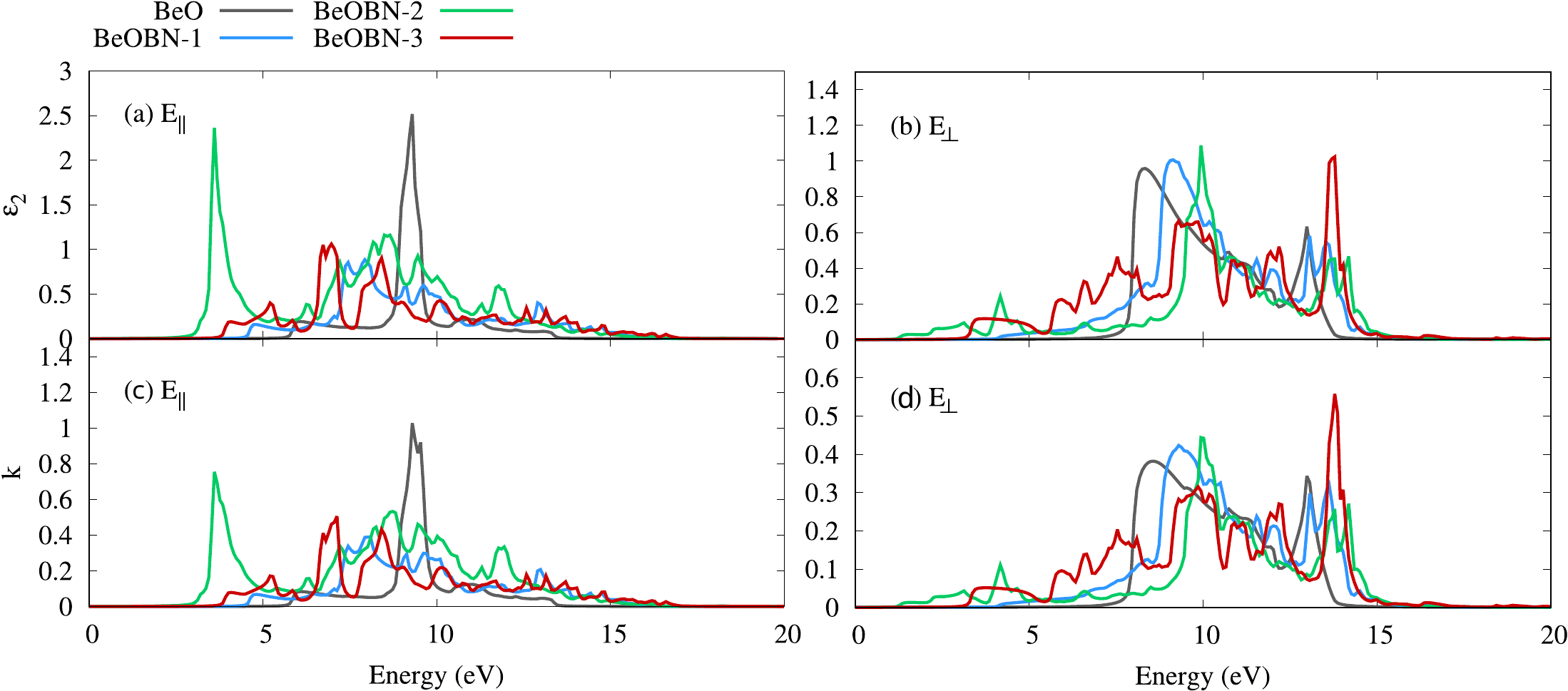}
	\caption{Imaginary part of the dielectric function, $\varepsilon_2$ (a) and (b), and excitation spectra, $k$ (c) and (d), for a parallel $E_{\rm \parallel}$ (left panel) and perpendicular $E_{\rm \perp}$ (right panel) polarizations of the electric field, respectively.}
	\label{fig04}
\end{figure*}

We first start with the $\varepsilon_2$, and $k$, which are shown in \fig{fig04} for 
parallel polarization, $E_{\rm \parallel}$ (a,c), and perpendicular polarization, $E_{\rm \perp}$ (b,d), of an electric field.
We see threshold energies for pure BeO monolayer to be $5.85$ and $7.81$~eV obtained from the $\varepsilon_2$ spectra for $E_{\rm \parallel}$ and $E_{\rm \perp}$, respectively. 
Strong peak in both $\varepsilon_2$, and $k$ are seen at $9.3$~eV for $E_{\rm \parallel}$ demonstrating a transition from the valence band maxima to the conduction band at the K point.
\lipsum[0]
\lipsum[0]
\begin{figure*}[htb]
	\centering
	\includegraphics[width=0.85\textwidth]{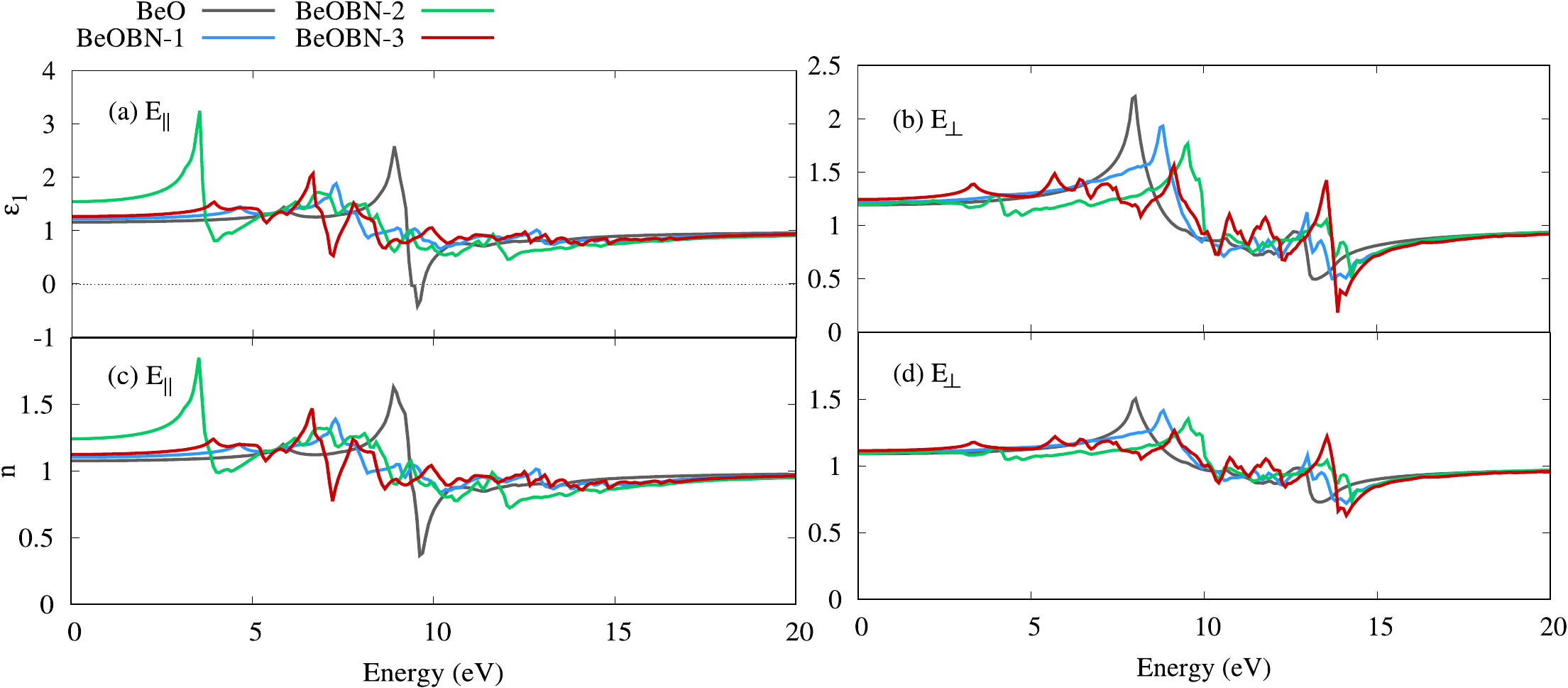}
	\caption{Real part of the dielectric function, $\varepsilon_1$ (a) and (b), and refractive index, $n$ (c) and (d), for a parallel $E_{\rm \parallel}$ (left panel) and perpendicular $E_{\rm \perp}$ (right panel) polarizations of the electric field, respectively.}
	\label{fig05}
\end{figure*}
The threshold energies and the peak positions tell us that the physical properties of pure BeO monolayer are different from the properties of BeO bulk in its wurtzite structure \cite{Valedbagi2014, Amrani_2007}, which is due to the Be-O hybridization. In a BeO monolayer each atom (Be or O atoms) has three bonds with the nearest neighbor, but in the BeO wurtzite structure each atom is bonded with four other atoms.

For the BeOBN monolayers, the threshold energies, and the intensity and the position of the intense peak are changed. The threshold energy is decreased to $4.14$, $1.62$, $3.33$~eV for BeOBN-1, BeOBN-2, and BeOBN-3 monolayers, respectively, which arise from the reduction of their band gaps in the $E_{\rm \parallel}$ direction. The intense peak is red shifted to a lower energy in the $E_{\rm \parallel}$, while it is blue shifted to a higher energy in the $E_{\rm \perp}$. 
The strong reduction in the band gap of BeOBN-2 induces a high intense peak at low energy range in both the $\varepsilon_2$ and the $k$ spectra at $3.6$~eV, which will be beneficial for optoelectronic devices working in the visible light region. In addition, the parallel bands of BeOBN monolayers lead to most of the transitions in the visible and the UV regions as has been previously reported \cite{JOHN2017307}.
We can therefore see several peaks in the $\varepsilon_2$ and the $k$ spectra of BeOBN monolayers in addition to the main intense peak. An interesting phenomena, the van Hove singularity, has been reported caused by the flat or the parallel bands leading to optical transitions in 2D materials \cite{JOHN2017307, ABDULLAH2021114644}.

The real part of the dielectric function and the refractive index are shown in \fig{fig05} for the 
$E_{\rm \parallel}$ (left panel) and $E_{\rm \perp}$ (right panel). The refractive index spectra follows almost the same characterstic as $\varepsilon_1$.
The $\varepsilon_1$, and $n$ have many significant variations with respect to the redistribution of the electron charge in the presence of the BN codopat atoms.
The sharp peak for pure BeO monolayer occurring at about $8.9$~eV becomes weaker and shifts towards lower energy ranges for BeOBN-1 and BeOBN-3 in the presence of $E_{\rm \parallel}$, while it gets stronger 
for BeOBN-2 appearing at a lower energy at $3.53$~eV.
In addition, the static dielectric constant, $\varepsilon_1(0)$, becomes a bit larger.
While, the BN codopant has different effects on $\varepsilon_1$ and $n$ in $E_{\rm \perp}$ direction. In these spectra, the main peak at $8.0$~eV for pure BeO becomes weaker and is
dislocated to higher energy ranges about $9.5$~eV for BeOBN monolayers.
These phenomena arise due to the variation of the in-plane bond nature in which the
$\sigma$ bonds get stronger especially for BeOBN-2 and a side-by-side overlapping is decreased.
The strong $\sigma$ bond of BeOBN-2 monolayer is caused by the strong interaction between the N and O atoms shown in \fig{fig01}(c). Thus the sharp peak
for the $E_{\rm \parallel}$ polarization gets stronger, while it is reduced for
$E_{\rm \perp}$.

The plasmon is the collective oscillations of the free electron density in a material. We can 
obtain the value for the plasmon frequency from $\varepsilon_1$. At the plasmon frequency, the value of $\varepsilon_1$ should be zero, and the value of $\varepsilon_2$ attains a maximum value.
So, one can see that the plasmon frequency for pure BeO monolayer is $9.3$~eV for $E_{\rm \parallel}$, while no plasmon mode is found for the $E_{\rm \perp}$. 
It should be noticed that the plasmon frequency vanishes for all three types of the BeOBN monolayers. 
The disappearance of the plasmon frequency is ascribed to the replacement of O or Be atoms by B and N atoms
in which an electron donation to the O atoms from the B and N atoms is seen.

Generally, the BN-codopant has anisotropic effects on the dielectric function spectrum for both  $E_{\rm \parallel}$ and $E_{\rm \perp}$ components. For example, the refractive index at the zero energy $n(0)$ is increased in the $E_{\rm \parallel}$, while it is almost unchanged in the $E_{\rm \perp}$ polarization.

\begin{figure}[htb]
	\centering
	\includegraphics[width=0.45\textwidth]{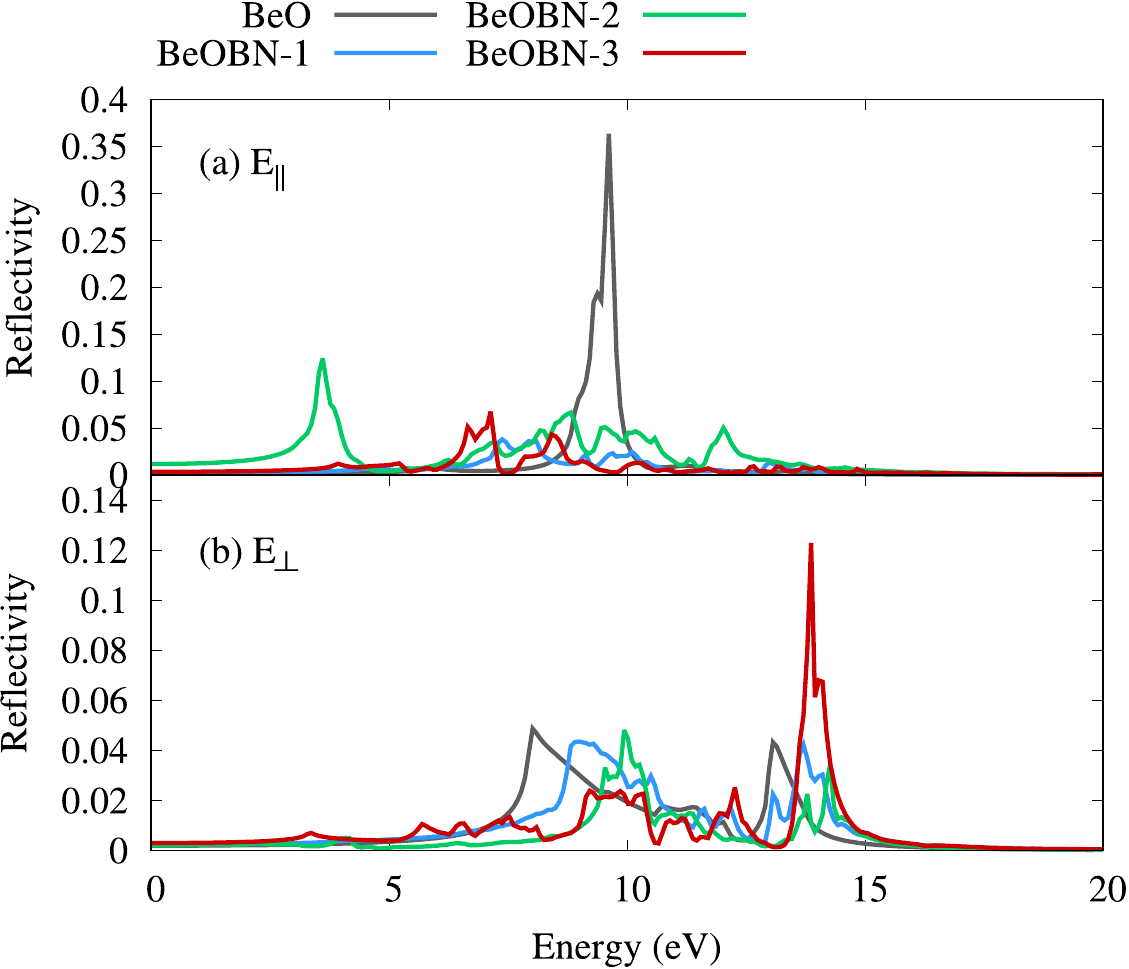}
	\caption{Reflectivity spectra for BeO and BeOBN monolayers in the case of parallel $E_{\rm \parallel}$ (a) and perpendicular $E_{\rm \perp}$ (b) polarizations of the electric field.}
	\label{fig06}
\end{figure}

To investigate the value of light reflectivity off the BeO and BeOBN monolayers, the reflectivity spectrum is studied based on the real and the imaginary part of the dielectric function \cite{JALILIAN2016120}. The reflectivity spectra is the effectiveness in reflecting radiation energy. 
The reflectivity spectra for BeO and BeOBN monolayers are shown in \fig{fig06} in the case of parallel $E_{\rm \parallel}$ (a) and perpendicular $E_{\rm \perp}$ (b) polarizations for the electric field. The reflectivity peaks correspond to the peaks of the real function seen in \fig{fig05}. The main strong peak seen for a BeO monolayer at $9.62$~eV is strongly decreased for the BeOBN monolayers indicating a less reflective behavior due to electron charge redistribution arising from the B and N atoms.
Furthermore, the reflection percentage in the visible range of the electromagnetic spectrum for $E_{\rm \parallel}$ component is about $12.5\%$ for BeOBN-2 monolayer, which indicates the transparency ratio is about $87.5\%$. The pure BeO and BeOBN monolayers become $100\%$ transparent after $15.0$~eV and $17.0$~eV for $E_{\rm \parallel}$, and $E_{\rm \perp}$, respectively. 

\begin{figure}[htb]
	\centering
	\includegraphics[width=0.45\textwidth]{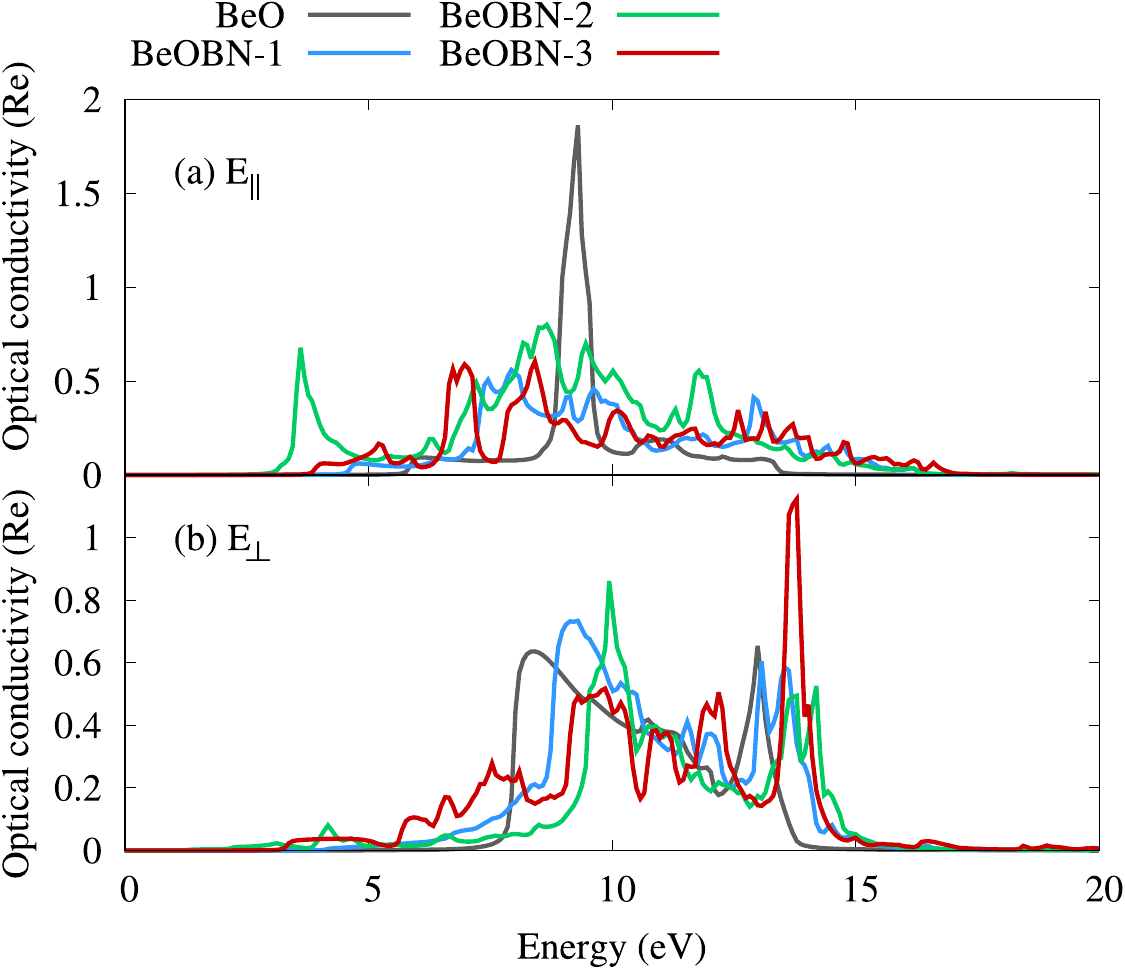}
	\caption{Optical conductivity (real part) for BeO and BeOBN monolayers in the case of parallel $E_{\rm \parallel}$ (a) and perpendicular $E_{\rm \perp}$ (b) polarizations of the electric field}
	\label{fig07}
\end{figure}

Another interesting physical behavior of the structures is their optical conductivity, which is presented in 
\fig{fig07} for parallel $E_{\rm \parallel}$ (a) and perpendicular $E_{\rm \perp}$ (b) polarizations of the electric field. The optical conductivity in $E_{\rm \parallel}$ is almost zero up to $5.6$, $4.15$, $1.62$, and $3.332$ for BeO, BeOBN-1, BeOBN-2, BeOBN-3 confirming the semiconducting behavior of these monolayers. 

In the case of $E_{\rm \perp}$, the optical conductivity of BeO is zero up to $7.8$~eV confirming the insulating behavior, while the optical conductivity is increased in the low energy ranges for the BeOBN monolayers. It becomes zero below $3.58$, $1.0$, and $2.2$~eV for BeOBN-1, BeOBN-2, and BeOBN-3, respectively. It reveals that the BeOBN monolayers have semiconducting properties in the presence of $E_{\rm \perp}$. In addition, the main strong peak for a BeO monolayer is red shifted to the lower energy for $E_{\rm \parallel}$ while it is blue shifted for $E_{\rm \perp}$. A strong peak at $3.6$~eV is found for BeOBN-2 arsing from the transition from valence band to the conduction band due to reduction of the band gap. 
In general, the optical response of the BeOBN monolayers at low energy ranges of our calculations indicates that these structures will be useful for optoelectronic devices in the visible light range of electromagnetic spectra.

\subsection{Thermoelectric properties}

Our next step is the study of the thermoelectric properties of our structures at low temperature ranges up to $200$~K. It has been shown that the electron and phonon temperatures are decoupled at the 
selected range of temperature \cite{PhysRevB.87.241411}. 
There have been several attempts to increase the figure of merit, $ZT$, which reflects the efficiency of thermoelectric structure. The $ZT$ is defined as \cite{nano9020218, B822664B}
\begin{equation}
	ZT = \frac{S^2 \sigma}{k} T,
\end{equation}\label{eq01}
where $S$ indicates the Seebeck coefficient, $\sigma$ is the electrical conductance, $k$ is the electronic thermal conductance, and $T$ is the temperature. In order to obtain a high $ZT$, we should have high values of $S$ and $\sigma$, and a low value of $k$ \cite{ABDULLAH2021413273}.  

\begin{figure}[htb]
	\centering
	\includegraphics[width=0.45\textwidth]{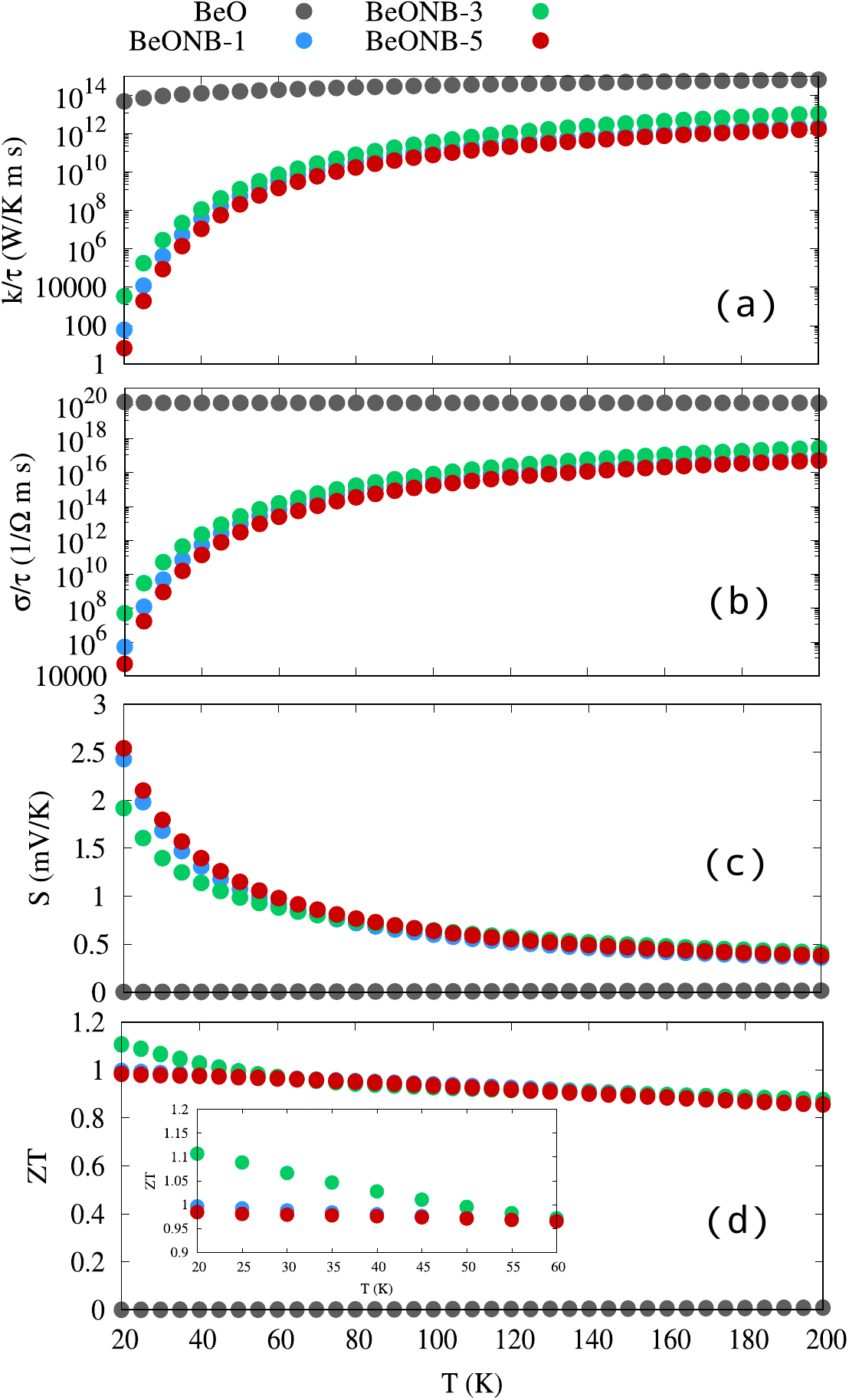}
	\caption{Electronic thermal conductivity, $k$ (a), electrical conductivity, $\sigma$ (b), 
	Seebeck coefficient, $S$ (c), and Figure of merit, $ZT$ (d), versus temperature for 
    BeO (gray), BeOBN-1 (blue), BeOBN-2 (green), and BeOBN-3 (red) monolayers.} 
	\label{fig08}
\end{figure}

The temperature-dependence of the electronic thermal conductivity (a), the electrical conductivity (b), 
the Seebeck coefficient (c), and the Figure of merit (d) are presented for
BeO (gray), BeOBN-1 (blue), BeOBN-2 (green), and BeOBN-3 (red) monolayers in \fig{fig08}.
We select the values with chemical potential near to the valence band
maxima because the BeO and BeOBN monolayers are p-type thermoelectric materials as 
can be clearly seen from their band structures.

However the electrical conductivity is decreased for the BeOBN monolayers, but the electronic thermal conductivity is also decreased over the selected range of temperature for all the
considered BeOBN monolayers, which is a good achievement towards increasing the figures of merit based on \eq{eq01}. On the other hand, the Seebeck coefficient is enhanced for the BeOBN monolayers. 
In order to understand the enhancement nature of Seebeck coefficient, we refer to the band structure of these monolayers, as the band structures are very important for understanding thermoelectric properties. 
For all the three BeOBN structures, the edges of the conduction bands and the valence bands are asymmetric. In addition, the band gap is also reduced especially for the BeOBN-2 monolayer. 
We also focus on the bands closest to the Fermi level because they are important to determine the transport properties. It can be seen that the degenerate states close to the Fermi energy in the valence band region are lifted for BeOBN forming asymmetry properties of the band structures.
Because of the aforementioned reasons, asymmetry properties, and the reduction of band gap, the Seebeck coefficient is enhanced for the BeOBN monolayes. As a results, the figure of merit is enhanced over the entire temperature range for the BeOBN monolayes. The enhancement of the thermoelectric properties is higher for the low temperature range from $20$ to $50$~K, especially for the BeOBN-2 monolayer (see inset figure), where the value of $ZT$ approaches 1.11. This is expected, because at a very low temperature the phonons are strongly frozen. 
Our results for the thermoelectric properties of the BeOBN monolayers may be important as the 
enhancement of the thermal properties of the BeO monolayer due to the BN codopant atoms will be useful for thermoelectric devices.

\section{Conclusion and Remarks}\label{Sec:Conclusion}

We have shown that the physical properties of BeO monolayers can be enhanced using BN-codopant atoms with different atomic configuration of the B and N atoms.
The electronic and the optical properties are calculated using density functional theory, and the 
thermoelectric properties, such as the thermal and electrical conductivity, the Seebeck coefficient, and the figure of merit are evaluated using the Boltzmann transport equation. It was shown that the B and N dopant atoms interact with the Be and O atoms in the hexagon structure of the BeO monolayer. The strong interaction between the O and the N atoms emerges as they have very close electronegativity. The band gap is thus reduced which converts the insulator properties of BeO monolayer to a semiconductor. The reduced band gap increases the efficiency of the thermoelectric characteristics via the ZT, and generates a high optical conductivity at the low energy.

\section{Acknowledgment}
This work was financially supported by the University of Sulaimani and 
the Research center of Komar University of Science and Technology. 
The computations were performed on resources provided by the Division of Computational 
Nanoscience at the University of Sulaimani.  
 
%\section{References}

%\bibliographystyle{elsarticle-num} 
%\bibliography{Ref_2.bib}

\end{document}